\documentclass[%
 reprint,
 amsmath,amssymb,
 aps,
]{revtex4-2}

\usepackage{graphicx}
\usepackage{dcolumn}
\usepackage{bm}

\usepackage{gensymb}
\usepackage{amsmath}

\usepackage{textcomp}
\usepackage{float}

\begin{document}

\preprint{APS/123-QED}

\title{Nonreciprocal reflection of mid-infrared light\\by highly doped InAs at low magnetic fields}

\author{Simo~Pajovic$^{1}$}
\email{pajovics@mit.edu}
\author{Yoichiro~Tsurimaki$^{2}$}
\author{Xin~Qian$^{3}$}
\author{Gang~Chen$^{1}$}
\author{Svetlana~V.~Boriskina$^{1}$}
\email{sborisk@mit.edu}

\affiliation{$^{1}$ Department of Mechanical Engineering, Massachusetts Institute of Technology, Cambridge, MA 02139, USA\looseness=-1}
\affiliation{$^{2}$ Department of Electrical Engineering, Stanford University, Stanford, CA 94305, USA\looseness=-1}
\affiliation{$^{3}$ School of Energy and Power Engineering, Huazhong University of Science and Technology, Wuhan 430074, China\looseness=-1}

\date{\today}

\begin{abstract}
We report an experimental observation of room-temperature nonreciprocal reflection of mid-infrared light from planar highly doped InAs surfaces at low magnetic fields ranging from 0.07 T to 0.16 T. Using ellipsometry, we demonstrate that the amplitude ratio and phase shift of reflected light are nonreciprocal in the Voigt configuration. We also demonstrate using Fourier-transform infrared spectroscopy that the nonreciprocal reflectance contrast (the difference in reflectance in opposite directions) increases with the magnitude of the magnetic field for \textit{p}-polarized light. Our work is a step toward the practical implementation of nonreciprocal thermal emitters and absorbers and applications such as remote magnetic field sensing.
\end{abstract}

\maketitle


\section{\label{sec:intro}Introduction}

Lorentz reciprocity states that a source and a detector of light can be freely interchanged without changing the outcome of the measurement by the detector \cite{lorentz1896original,landaulifshitz,asadchy2020}. Reciprocity is useful for applications such as telecommunication, where it implies that an antenna that is a good receiver is a good transmitter as well. However, reciprocity assumes that light propagates through media whose optical properties (i.e., permittivity, permeability, and conductivity) are (1) linear, (2) time-invariant, and (3) described by symmetric tensors. In principle, breaking one or more of these assumptions would break Lorentz reciprocity, and the medium could support ``nonreciprocal'' electromagnetic modes, via the electromagnetic wave equation.

In recent years, the prospect of electromagnetic nonreciprocity has become particularly exciting because it implies that Kirchhoff's law of thermal radiation---the equality of spectral directional emissivity and absorptivity \cite{kirchhoff1860original,siegelhowell}---can be violated \cite{remer1984,snyder1998,zhu2014,tsurimaki2020}. In particular, it can be shown that when nonreciprocity is present, emissivity $\epsilon$, absorptivity $\alpha$, and reflectance $R$ are related by the generalized Kirchhoff's law of thermal radiation:

\begin{equation} \label{eq:kirchhoff}
    \epsilon_{\ell}(\omega,\theta)-\alpha_{\ell}(\omega,\theta)=R_{\ell}(\omega,+\theta) - R_{\ell}(\omega,-\theta),
\end{equation}

\noindent where $\omega$ is the angular frequency, $\theta$ is the polar angle of incidence, and $\ell$ is the polarization of light. Eq. (\ref{eq:kirchhoff}) assumes specular reflection \cite{remer1984,zhu2014}; otherwise, Eq. (\ref{eq:kirchhoff}) can be written in terms of the bidirectional reflectance distribution function, or BRDF \cite{snyder1998,tsurimaki2020}. If the left-hand side of Eq. (\ref{eq:kirchhoff}) is nonzero, it implies that Kirchhoff's law of thermal radiation has been violated, and the difference $R_{\ell}(\omega,+\theta) - R_{\ell}(\omega,-\theta)$, or nonreciprocal reflectance contrast, is a metric of the inequality of spectral directional emissivity and absorptivity.

Nonreciprocity and the violation of Kirchhoff's law of thermal radiation have implications for the energy efficiency of devices that emit and absorb thermal radiation such as solar thermal collectors and thermophotovoltaics since they no longer have to emit as much as they absorb through any given spectral directional channel \cite{green2012,park2021}. In addition, nonreciprocity allows for unprecedented control of thermal radiation, such as directionally asymmetric emission and absorption. Technologies including thermal imaging, tailored thermal signatures, and radiative cooling can benefit from this level of control. For most of these applications, the emitter or absorber is at room temperature, so as a result of Wien's law \cite{wien1896original,siegelhowell}, candidate materials should support nonreciprocal electromagnetic modes in the mid-infrared (mid-IR).

There are a number of different physical mechanisms that can enable nonreciprocity. Nonlinear media can work, but the intensity of emitted or absorbed light needs to be high (which is not the case for room-temperature thermal radiation) or some other high power input is needed to drive nonlinear effects \cite{fan2012nonlinear,mahmoud2015nonlinear,shi2015nonlinear,lawrence2018nonlinear}. Space and/or time modulation can work as well and have been experimentally demonstrated at lower frequencies than the mid-IR \cite{yu2009spacetime,lira2012spacetime,hadad2016spacetime,taravati2016spacetime,cardin2020spacetime}. Time modulation has been theoretically predicted to have fundamentally interesting and potentially useful consequences for thermal radiation, such as coherence and optical refrigeration \cite{renwen2023,renwen2024}. In particular, spatiotemporally modulated metasurfaces are tailorable and have been shown to have the potential to become platforms for mid-IR nonreciprocity \cite{povinelli2022,ghanekar2022}, but the required frequency modulation ($\gtrsim$ GHz) is difficult to achieve in practice.  A direct current (DC) magnetic field, which breaks assumption (3) of Lorentz reciprocity by inducing antisymmetric components of the dielectric tensor, may be the most well-known approach to mid-IR nonreciprocity.

Within the scope of nonreciprocity enabled by magnetic fields, in the literature, there has been an emphasis on two classes of materials: highly doped semiconductors such as InSb and InAs, which support free-carrier magneto-optic effects \cite{palik1967,zvezdinkotov}, and magnetic Weyl semimetals such as $\textrm{Co}_{2}\textrm{MnGa}$ \cite{belopolski2019co2mnga} and $\textrm{Co}_{3}\textrm{Sn}_{2}\textrm{S}_{2}$ \cite{morali2019co3sn2s2} (although only the former has a Curie temperature above room temperature). Theoretical research on these candidate materials shows a great deal of promise: in combination with design strategies from the field of nanophotonics, such as multilayer structures \cite{wu2022multilayer1,wu2022multilayer2,wang2023,gold2024}, gratings \cite{zhu2014,zhao2019,wu2022grating,wu2023grating}, metamaterials \cite{li2024}, and prisms \cite{wu2021}, InAs-based absorbers have been theoretically predicted to be able to ``totally'' violate Kirchhoff's law, i.e., nearly 100$\%$ emissivity and 0$\%$ absorptivity at a given frequency and angle of incidence \cite{zhu2014,zhao2019,wang2023}. However, the magnetic fields required can be large, up to $\sim 1$ T, comparable to that generated by a magnetic resonance imaging (MRI) scanner. On the other hand, magnetic Weyl semimetals are theoretically predicted to support stronger nonreciprocity than InAs without the need for applied magnetic fields. This is because in the electronic band structure of these materials, the Berry curvature flux between Weyl nodes of opposite chirality behaves like a pseudo-magnetic field in momentum space \cite{yanfelser2017review}, giving rise to the anomalous Hall effect and large off-diagonal components of the dielectric tensor compared to the diagonal components \cite{hofmann2016,zhao2020,tsurimaki2020,pajovic2020}.

Despite the vast body of work on theoretically predicting and optimizing nonreciprocal emission and absorption in the mid-IR, experiments are scarce \cite{yang2024}. Although magneto-optic effects such as the Faraday effect have been explored in the context of magnetic Weyl semimetals \cite{han2022}, to date, there have been no direct observations of mid-IR nonreciprocity, e.g., measurements of directionally asymmetric (``nonreciprocal'') reflectance $R(\omega,+\theta)\neq R(\omega,-\theta)$, where $\omega$ is the angular frequency and $\theta$ is the polar angle of incidence. Historically, InAs and InSb have been the platforms of choice for experimental demonstrations of magnetic-field-enabled mid-IR nonreciprocity.  Although some experiments have succeeded in directly observing nonreciprocity in lightly doped InAs and InSb \cite{remer1984,heyman2001,chochol2017}, these results have limited applications to thermal radiation because of cryogenic temperatures and/or low frequencies (on the order of $10^0$ THz instead of $10^1$ THz). More recent experiments that used lightly doped InSb in combination with metasurfaces have similar limitations \cite{keshock2020,peng2022,peng2023}, although the Verdet constant (which is related to the off-diagonal components of the dielectric tensor) of highly doped InSb has been measured in the mid-IR \cite{peard2021}. In the past few years, highly doped InAs has been used to experimentally demonstrate nonreciprocal absorption in the mid-IR via guided-mode resonances \cite{shayegan2022} and epsilon-near-zero (ENZ) modes \cite{liu2023}. In fact, the violation of Kirchhoff's law was directly observed in these systems, i.e., experimental values of spectral directional emissivity and absorptivity were not equal \cite{shayegan2023,shayegan2024}. However, similar to many theoretical predictions, these experiments required large magnetic fields ranging from 0.5 T to 1.5 T, which may not be practical or integrable. Strong nonreciprocity is theoretically possible for these materials at magnetic fields as low as 0.3 T (to be discussed, see also  \cite{zhao2019}, for example), but experiments in this regime are lacking. Given this limitation and the small number of experiments that have been conducted on highly doped semiconductors in the context of mid-IR nonreciprocity, more work is needed to fully understand their capabilities, especially at low magnetic fields.

To help fill these gaps, we investigated the mid-IR nonreciprocal optical response of planar highly doped InAs surfaces at magnetic fields ranging from 0.07 T to 0.16 T, which were generated using off-the-shelf neodymium magnets. Using ellipsometry, we experimentally observed that the amplitude ratio and phase difference between \textit{s}- and \textit{p}-polarized light, $\Psi$ and $\Delta$, are nonreciprocal in the Voigt configuration (magnetic field pointing normal to the plane of incidence). In addition, we showed that it is possible to fit the dielectric tensor of highly doped InAs to measured $\Psi$ and $\Delta$ spectra, without the need for time-consuming Mueller matrix measurements. Then, using Fourier-transform infrared spectroscopy (FTIR), we experimentally observed that the reflectance of \textit{p}-polarized light is nonreciprocal and that the nonreciprocal reflectance contrast, defined as the difference in reflectance in opposite directions $+\theta$ and $-\theta$, increases with increasing magnetic field. Our work demonstrates that measurable nonreciprocal reflectance contrasts can be achieved at low magnetic fields, at thermally relevant wavelengths in the mid-IR. This is an important step toward making nonreciprocal thermal emitters and absorbers practical, which has the potential to improve technologies such as solar energy harvesting, thermal imaging, thermal management, and radiative cooling.

In what follows, we will review the theory of nonreciprocal reflection due to free-carrier magneto-optic effects in highly doped semiconductors. Then, we will describe our experimental setups for ellipsometry and FTIR and the magnetization schemes we designed and implemented. Finally, we will discuss the results of our experiments and share our outlook on the field of nonreciprocal thermal radiation, with an emphasis on the potential of highly doped III-V semiconductors in the same family as InAs.

\section{\label{sec:theory}Theory}

\subsection{Material model and reflection coefficients}

Generally speaking, free-carrier magneto-optic effects can be modeled using the gyrotropic Drude-Lorentz model \cite{seeger}. In this extension of the Drude model, the Lorentz force is included, which gives rise to off-diagonal components of the dielectric tensor. If the magnetic field points in the $y$-direction, i.e.,  $\textbf{B}=B\hat{\textbf{y}}$, the dielectric tensor can be written in the form

\begin{equation} \label{eq:epsilon}
\overline{\overline{\epsilon}}=
\begin{bmatrix}
\epsilon_T & 0 & +ig\\
0 & \epsilon_L & 0\\
-ig & 0 & \epsilon_T
\end{bmatrix},
\end{equation}

\noindent where $\epsilon_T$ and $\epsilon_L$ are the transverse and longitudinal components of $\overline{\overline{\epsilon}}$ and $g$ is the gyration vector. These depend on $\omega$ and $B$ as follows:

\begin{align}
    \epsilon_{T}(\omega,B) &= \epsilon_{\infty} - \frac{\omega_{p}^{2}(\omega+i\gamma)}{\omega\left[(\omega+i\gamma)^{2}-\omega_{c}^{2}\right]}, \label{eq:transverse}\\
    \epsilon_{L}(\omega) &= \epsilon_{\infty} - \frac{\omega_{p}^{2}}{\omega(\omega+i\gamma)^{2}}, \label{eq:longitudinal}\\
    g(\omega,B) &= \frac{\omega_{p}^{2}\omega_{c}}{\omega\left[(\omega+i\gamma)^{2}-\omega_{c}^{2}\right]}. \label{eq:gyration}
\end{align}

\noindent In Eqs. (\ref{eq:transverse})--(\ref{eq:gyration}), $\epsilon_{\infty}$ is the high-frequency dielectric constant, $\omega_p = \sqrt{ne^2/\epsilon_{0}m^*}$ is the plasma frequency, $\omega_c = eB/m^*$ is the cyclotron frequency, and $\gamma$ is the damping constant. Furthermore, $n$ is the carrier concentration, $m^*$ is the electronic effective mass, and $e$ and $\epsilon_0$ are the elementary charge and permittivity of free space, respectively. Equations (\ref{eq:epsilon})--(\ref{eq:gyration}) are applicable to magnetoplasmas and materials that support free-carrier magneto-optic effects, such as highly doped InAs, InSb, and other $n$-type semiconductors.

\begin{figure*}
\centering\includegraphics[scale=0.75]{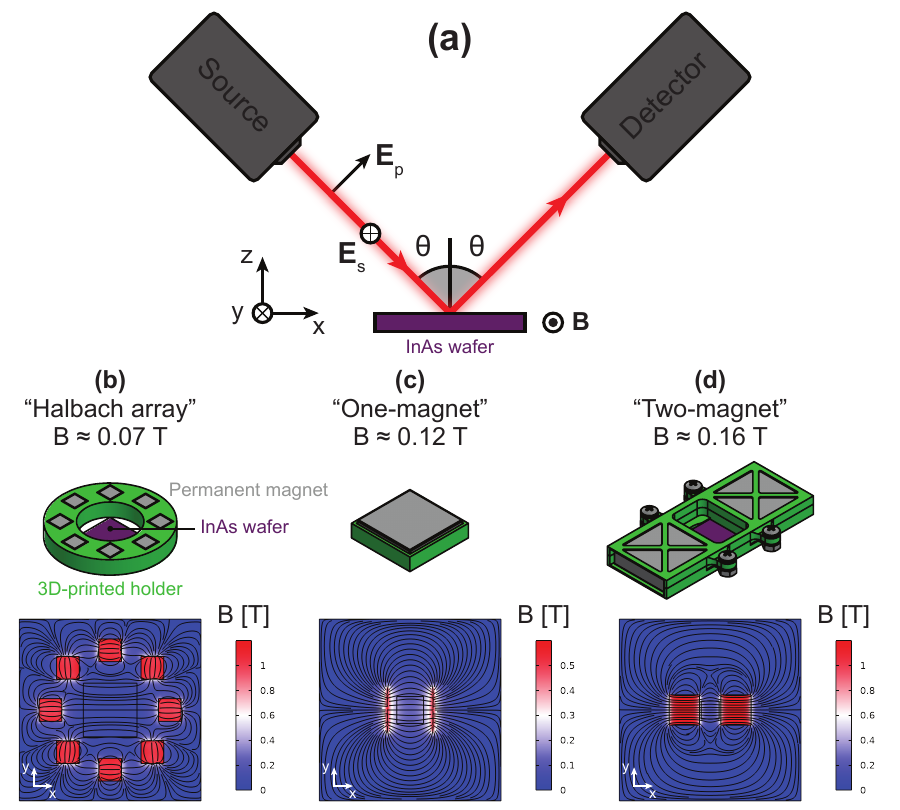}
\caption{Mid-infrared spectroscopy of highly doped InAs at low magnetic fields. (a) Schematic of the experimental setup, applicable to both ellipsometry and Fourier-transform infrared spectroscopy (FTIR). Light is incident on the InAs wafer at polar angle of incidence $\theta$ and can be \textit{s}- or \textit{p}-polarized, indicated by the electric field vectors $\textbf{E}_s$ and $\textbf{E}_{p}$, respectively. The InAs wafer is magnetized in the $y$-direction, normal to the plane of incidence (the Voigt configuration), in which case there is no rotation of the plane of polarization. There are three magnetization schemes providing three different magnetic fields $B$; schematics of the geometry and the magnetic field distributions in the lateral ($xy$-) plane are shown for each one. (a) Halbach array, $B\approx0.07$ T. (b) One-magnet, $B\approx0.12$ T. (c) Two-magnet, $B\approx0.16$ T.}
\label{fig:setup}
\end{figure*}

Now consider the interface between air (having dielectric constant 1) and a magneto-optic material having a dielectric tensor of the form of Eq. (\ref{eq:epsilon}). It is known from symmetry arguments \cite{remer1984,camley1987,figotin2001} that nonreciprocal reflectance is maximized in the Voigt configuration, where $\textbf{B}$ is perpendicular to the plane of incidence or wavevector of light. The other limiting case is the Faraday configuration, where $\textbf{B}$ is parallel to the plane of incidence and reflection is reciprocal. In Eq. (\ref{eq:epsilon}), $\textbf{B}$ points in the $y$-direction, so light propagates in the $xz$-plane, as illustrated in Fig. \ref{fig:setup}(a). In the Voigt configuration, the electromagnetic modes supported by the magneto-optic material are \textit{p}- and \textit{s}-polarized, which greatly simplifies the analysis since there is no rotation of the plane of polarization upon reflection. It can be shown that the \textit{p}- and \textit{s}-polarized reflection coefficients are

\begin{alignat}{2}
    r_p &\equiv \frac{\textbf{E}_{p}^{r}\cdot\hat{\textbf{x}}}{\textbf{E}_{p}^{i}\cdot\hat{\textbf{x}}} & &= \frac{\left(k_0^2 - k_{z}^{i}k_{z,p}^{t}\right)\epsilon_{T} + (igk_{z}^{i}-k_{x})k_{x}}{\left(k_0^2 + k_{z}^{i}k_{z,p}^{t}\right)\epsilon_{T} - (igk_{z}^{i}+k_{x})k_{x}}, \label{eq:rp}\\
    r_s &\equiv \frac{\textbf{E}_{s}^{r}}{\textbf{E}_{s}^{i}} & &= \frac{k_{z}^{i} - k_{z,s}^{t}}{k_{z}^{i} + k_{z,s}^{t}}, \label{eq:rs}
\end{alignat}

\noindent where the superscripts $i$, $r$, and $t$ indicate incident, reflected and transmitted light, respectively \cite{remer1984,khanikaev2009,pajovic2020}. In Eqs. (\ref{eq:rp}) and (\ref{eq:rs}), $k_0=\omega/c$, where $c$ is the speed of light in free space, $k_{x}=k_{0}\sin{\theta}$ is the $x$-component of the wavevector, and $k_{z}^{i}=\sqrt{k_{0}^{2}-k_{x}^{2}}$ is the $z$-component of the wavevector in air. In the magneto-optic material, the $z$-component of the wavevector is polarization-dependent (i.e., it is birefringent):

\begin{align}
    k_{z,p}^{t} &= \sqrt{\left(\epsilon_{T}-\frac{g^{2}}{\epsilon_{T}}\right)k_{0}^{2}-k_{x}^{2}}, \label{eq:kzp}\\
    k_{z,s}^{t} &= \sqrt{\epsilon_{L}k_{0}^{2}-k_{x}^{2}}. \label{eq:kzs}
\end{align}

\noindent Equations (\ref{eq:rp})--(\ref{eq:kzs}) assume that the relative permeabilities of both materials is 1, which is a reasonable assumption at optical frequencies \cite{landaulifshitz,pershan1967}. Note that only \textit{p}-polarized light is nonreciprocal since $r_{p}$ has terms that are linear in $k_{x}$, such that $r_{p}(+k_x)\neq r_{p}(-k_x)$. The linear coefficient in this case is $g$, which is nothing more than the $B$-dependent off-diagonal component of the antisymmetric dielectric tensor $\overline{\overline{\epsilon}}$. If $B=0$, these terms vanish, the dielectric tensor becomes symmetric, and $r_{p}$ becomes reciprocal. On the other hand, \textit{s}-polarized light is always reciprocal because its dispersion relation (Eq. (\ref{eq:kzs})) does not depend on $g$; it behaves like light in an isotropic material having dielectric constant $\epsilon_L$.

\subsection{Nonreciprocal measureables}

Nonreciprocity can manifest itself in a number of different measureables that are related to the reflection coefficients. In FTIR, the reflectance $R_{\ell}=|r_\ell|^2$ ($\ell=p,s$) is measured. If the sample is sufficiently optically thick such that there is no transmission, then the absorptivity can be calculated as $\alpha_{\ell}=1-R_{\ell}$. Information about thermal emission properties of the material can be inferred from the Onsager-Casimir reciprocal relations \cite{onsagercasimir} (or adjoint Kirchhoff's law \cite{guo2022}):

\begin{align}
    \alpha_{\ell}(\omega,+\theta,B)&=\varepsilon_{\ell}(\omega,-\theta,B),\label{eq:onsagertheta}\\
    \alpha_{\ell}(\omega,\theta,+B)&=\varepsilon_{\ell}(\omega,\theta,-B),\label{eq:onsagerb}
\end{align}

\noindent where $\varepsilon_{\ell}$ is the emissivity. In addition, Eqs. (\ref{eq:onsagertheta}) and (\ref{eq:onsagerb}) imply that flipping the sign of $\theta$ is equivalent to flipping the sign of $B$, i.e., $R_{\ell}(\omega,-\theta,B)=R_{\ell}(\omega,\theta,-B)$. This is extremely useful in experiments since it is usually easier to reverse the direction of $\textbf{B}$ (i.e., from $+\hat{\textbf{y}}$ to $-\hat{\textbf{y}}$) than it is to send light backwards from the detector to the source.

In ellipsometry, the ratio of $r_p$ to $r_s$---a complex number---is measured in terms of its amplitude and phase. More specifically, if

\begin{equation} \label{eq:ellipsometry}
\frac{r_p}{r_s} \equiv \tan{\Psi}e^{i\Delta},
\end{equation}

\noindent the ellipsometer measures $\Psi=\tan^{-1}(|r_{p}/r_{s}|)$ and $\Delta=\angle(r_{p}/r_{s})$, the amplitude ratio and phase difference between \textit{p}- and \textit{s}-polarized light, respectively \cite{fujiwara}. Since ellipsometry essentially measures a ratio, it can be highly sensitive and less susceptible to perturbations such as scattering and low intensity since they affect both the numerator and denominator, thus ``canceling out.'' For this reason, ellipsometry could be a more unambiguous way to experimentally probe nonreciprocity than FTIR.

\section{\label{sec:exp}Experiments}

Having reviewed the theory of nonreciprocal reflection, we now proceed to describing our experimental results. Figure \ref{fig:setup}(a) illustrates the experimental setup in both ellipsometry and FTIR (treating the source and detector of each one as black boxes, although they are different). Polarized light ($\textbf{E}_p$ or $\textbf{E}_s$ in Fig. \ref{fig:setup}(a)) is sent from the source toward a magnetized InAs wafer at polar angle of incidence $\theta$. The magnetic field $\textbf{B}$ points in the $y$-direction, which means that the entire system is in the Voigt configuration. The incident light is specularly reflected at the interface between air and the InAs wafer and collected by the detector. The ellipsometer we used was a J. A. Woollam Co. IR-VASE\textregistered\ Mark II, which allowed us to automatically control $\theta$ through the included software (WVASE\textregistered). The FTIR we used was a Thermo Scientific\texttrademark\ Nicolet\texttrademark\ iS50 FTIR Spectrometer. In this case, a PIKE Technologies VeeMAX III was used to manually control $\theta$, in conjunction with a ZnSe polarizer inserted into the accessory to manually control polarization.

To control $B$, we designed three magnetization schemes using off-the-shelf neodymium magnets (K\&J Magnetics, Inc.), which are illustrated in Figs. \ref{fig:setup}(b)--(d). Each magnetization scheme consists of an array of neodymium magnets held in place using a 3D printed holder and was designed to produce a magnetic field that is uniform in magnitude and direction across the InAs wafer. This can be seen in the magnetic field distribution plotted below each magnetization scheme in Figs. \ref{fig:setup}(b)--(d). These were calculated using COMSOL Multiphysics\textregistered. The ``Halbach array'' magnetization scheme produces a magnetic field of $B\approx0.07$ T, as shown in Fig. \ref{fig:setup}(b). ``One-magnet'' produces $B\approx 0.12$ T (Fig. \ref{fig:setup}(c)), and ``two-magnet'' produces $B\approx 0.16$ T (Fig. \ref{fig:setup}(d)). The calculated values of $B$ were experimentally validated by measuring the magnetic field as a function of position using a handheld teslameter. Details of the experimental validation can be found in Fig. \ref{fig:magnetization} in the Supplementary Material. Including $B=0$ T, this gave us the ability to measure four different levels of $B$.

Finally, we measured two InAs wafers, both of which were purchased from MSE Supplies LLC. Each sample was a square of side length 15 mm and 0.5 mm thick. We found that the doping levels of the samples were slightly different, which led to different strengths of nonreciprocity, to be discussed in the following sections.

\section{\label{sec:disc}Results and discussion}

\subsection{\label{sec:ellips}Ellipsometry}

\begin{figure*}
\centering\includegraphics{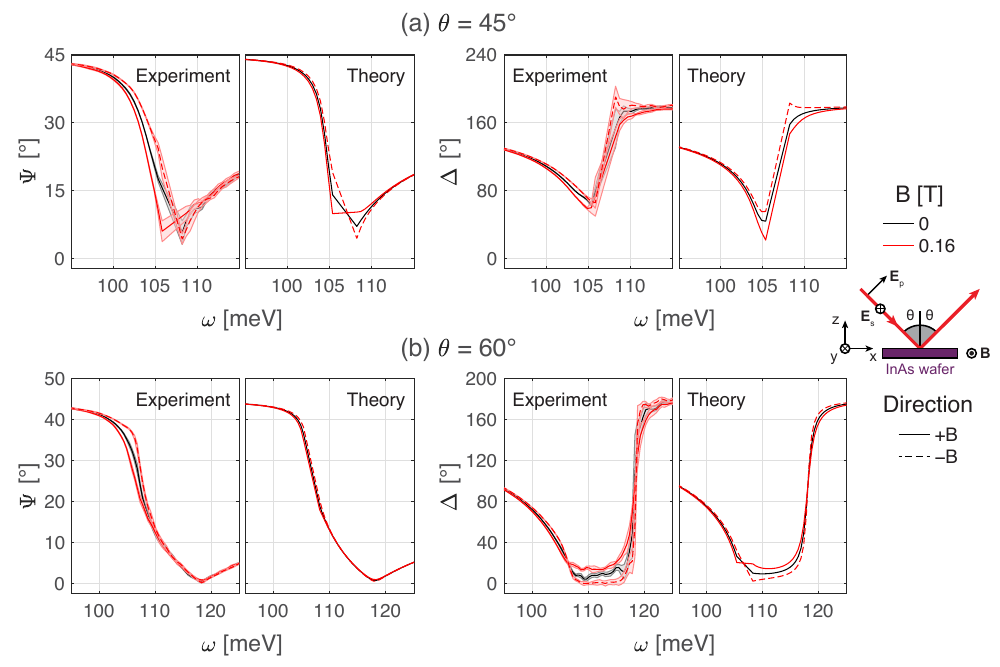}
\caption{Nonreciprocal reflection of mid-infrared light by highly doped InAs in a magnetic field, experimentally observed via ellipsometry. (a) Polar angle of incidence $\theta=45\degree$. Amplitude ratio $\Psi=\tan^{-1}(|r_{p}/r_{s}|)$ (left two columns) and phase difference $\Delta=\angle(r_{p}/r_{s})$ (right two columns) between \textit{s}- and \textit{p}-polarized light. Experimental and theoretical values, obtained by fitting the gyrotropic Drude-Lorentz model described in Sec. \ref{sec:theory}, are indicated. Black and red correspond to magnetic field $B$ values of 0 T and 0.16 T, respectively. Solid and dashed lines correspond to $B$ pointing in the $+y$- or $-y$-direction, respectively (equivalent to changing the sign of $\theta$). The shaded areas in the experimental values are the standard deviation ($N=50$). (b) Same as (a), but $\theta=60\degree$. The gyrotropic Drude-Lorentz model parameters were fitted using the data for both $\theta$ values. The inset schematic shows the coordinate system, the $B$ vector, $\theta$, and \textit{s}- and \textit{p}-polarized electric field vectors.}
\label{fig:ellips}
\end{figure*}

Figure \ref{fig:ellips} shows the results we obtained using ellipsometry. We measured $\Psi$ and $\Delta$ at two $\theta$-values: 45\degree and 60\degree, shown in Figs. \ref{fig:ellips}(a) and (b), respectively. In each panel, two values of $B$ are shown: 0 T and 0.16 T (two-magnet), corresponding to black and red lines. Data for $B=0.12$ T can be found in Fig. \ref{fig:addellips} in the Supplementary Material. Solid and dotted lines represent $B$ pointing in the $+y$- and $-y$-directions, respectively, which is equivalent to flipping the sign of $\theta$ via Eqs. (\ref{eq:onsagertheta})--(\ref{eq:onsagerb}). Experimental and theoretical $\Psi$ and $\Delta$ spectra experimental are shown. In the experimental values, the shaded areas represent the standard deviation of $N=50$ scans.

Since $\Psi$ and $\Delta$ can be written as closed-form expressions using Eqs. (\ref{eq:rp})--(\ref{eq:kzs}), the data can be fitted to the gyrotropic Drude-Lorentz model, the fit parameters being $\epsilon_\infty$, $n$, $m^*$, and $\gamma$. Typically, to fit the dielectric tensor of an anisotropic material, $\Psi$ and $\Delta$ measurements are not sufficient. Instead, the Mueller matrix is measured, which, although a standard procedure the vast majority of ellipsometers are capable of, can be extremely time-consuming---on the order of several hours per spectrum, and more than one spectrum, i.e., as a function of $\theta$, is usually needed for a good fit. By taking advantage of the symmetries of the Voigt configuration and the resulting simplicity of the equations for $r_p$ and $r_s$, we circumvented the need for Mueller matrix measurements and fitted the dielectric tensor of the sample using $\Psi$ and $\Delta$ as if it were isotropic. The results of the fit were $\epsilon_{\infty}=12.0$, $n=6.58\times10^{18}\ \textrm{cm}^{-3}$, $m^{*}=0.0724m_e$, and $\gamma=1.50$ meV. Hall effect measurements showed good agreement with the fitted value of $n$ ($<10\%$ error). These values are comparable to those reported in the literature for similar experiments on mid-IR nonreciprocity \cite{shayegan2022,shayegan2023,liu2023,shayegan2024}.

Generally speaking, there is good agreement between experimental and theoretical values of $\Psi$ and $\Delta$. There are two signatures of nonreciprocal reflection in the results: (1) $\Psi(\omega,\theta,+B)\neq\Psi(\omega,\theta,-B)$ and likewise for $\Delta$, and (2) the contrast between measurements at $+B$ and $-B$ increases with the magnitude of $B$, in an approximately monotonic way. These signatures are clearer for $\theta=60\degree$ (Fig. \ref{fig:ellips}(b)), especially when looking at $\Delta$ in the band 105--115 meV. The reason is that $\epsilon_{T}$ (Eq. (\ref{eq:transverse})) is near zero in this band \cite{liu2023}. In Eq. (\ref{eq:rp}), the $(k_{0}^{2}\pm k_{z}^{i}k_{z,p}^{t})\epsilon_{T}$ terms are always reciprocal because they do not linearly depend on $k_x$, whereas the $(igk_{z}^{i}\pm k_{x})k_{x}$ terms are always nonreciprocal as long as $g\neq 0$. In the limit as $\epsilon_{T}\rightarrow 0$, it can be shown that the reciprocal terms vanish and $r_{p}=(k_{x}-igk_{z}^{i})/(k_{x}+igk_{z}^{i})$, which only depends on $g$ and the frequency and direction of the incident light.

If $\Psi$ and $\Delta$ are calculated using a finer spectral resolution, there are a number of sharp features present in the theoretical values of $\Psi$ and $\Delta$ that are not apparent in the experimental values. This is shown in Fig. \ref{fig:theoellips} in the Supplementary Material. However, this could be because of the spectral resolution of the ellipsometer, which was 8 $\textrm{cm}^{-1}$ ($\approx1$ meV). A finer spectral resolution may have been able to resolve the sharp features, but this would have made the measurements much more time-consuming. On the other hand, it is worth noting that the sharp features could be used as spectral signatures for remote magnetic field sensing because (1) they are only present when $B\neq0$, (2) their bandwidth depends on the magnitude of $B$, and (3) their frequency (red- or blueshifted) depends on the direction of $B$. These trends can be seen in Fig. \ref{fig:ellips}(a) but not in Fig. \ref{fig:ellips}(b), which means that the sharp features are sensitive to $\theta$ as well. Since their bandwidth, frequency, and very presence are sensitive to $B$ and $\theta$, in principle, precise measurement of these spectral signatures could be used to remotely sense both the magnitude and the direction of a magnetic field using mid-IR light.

Ellipsometry has given us insight into how sensitive mid-IR nonreciprocity can be. Although nonreciprocity is sometimes thought of as ``weak'' in the sense that it is small effect that is difficult to measure and make practical use of and there are few candidate materials, our results suggest that experimental techniques such as ellipsometry can be highly sensitive to nonreciprocity, even at magnetic fields an order of magnitude lower than those generated in prior work ($\sim10^{-1}$ T compared to $\sim10^{0}$ T).

\subsection{\label{sec:ftir}FTIR}

\begin{figure*}
\centering\includegraphics{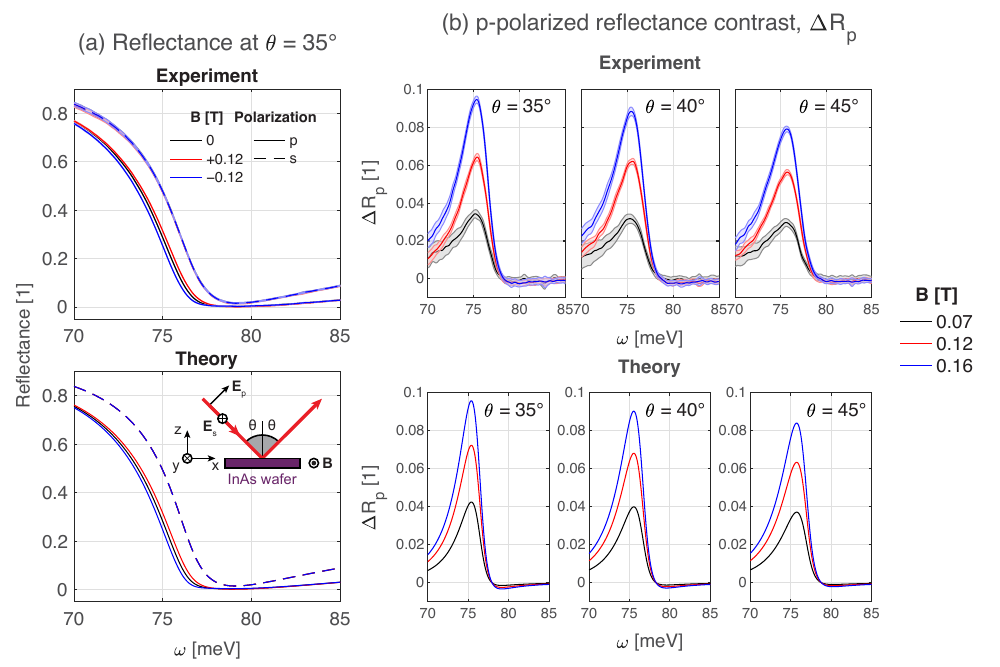}
\caption{Nonreciprocal reflectance of highly doped InAs at low magnetic fields. (a) Spectral reflectance $R_{\ell}(\omega,\theta,B)$ ($\ell=s,p$) at polar angle of incidence $\theta=35\degree$ as a function of magnetic field $B$ and polarization of light. Experimental (top) and theoretical (bottom) reflectance spectra, obtained by fitting the gyrotropic Drude-Lorentz model (note that this sample was different from the one in Fig. \ref{fig:ellips}). The black lines correspond to $B=0$ T; red $+0.12$ T (i.e., pointing in the $+y$-direction); and blue $-0.12$ T ($-y$-direction). Solid and dashed lines correspond to \textit{p}- and \textit{s}-polarization, respectively. The shaded areas in the experimental values are the standard deviation ($N=16$). The inset schematic shows the coordinate system, the $B$ vector, $\theta$, and \textit{s}- and \textit{p}-polarized electric field vectors. (b) Nonreciprocal reflectance contrast of \textit{p}-polarized light, $\Delta R_{p} \equiv R_{p}(\omega,\theta,+B)-R_{p}(\omega,\theta,-B)$. Experimental (top row) and (bottom row) theoretical $\Delta R_p$ at different $\theta$ values. The black lines correspond to $B=0.07$ T; red 0.12 T; and blue 0.16 T. The nonreciprocal reflectance contrast of \textit{s}-polarized light is experimentally and theoretically confirmed to be zero (see Fig. \ref{fig:spolnonrecip} in the Supplementary Material).}
\label{fig:ftir}
\end{figure*}

Figure \ref{fig:ftir} shows the results we obtained using FTIR. Figure \ref{fig:ftir}(a) shows the experimental and theoretical $R_p$ and $R_s$ spectra, using $\theta=35\degree$ as an example. Similar to ellipsometry, the data can be fitted to the gyrotropic Drude-Lorentz model using Eqs. (\ref{eq:rp})--(\ref{eq:kzs}) since they are closed-form expressions. The doping level of the sample we used in these experiments was different from the one used in the previous section. We found that $\epsilon_{\infty}=12.3$, $n=2.08\times10^{18}\ \textrm{cm}^{-3}$, $m^{*}=0.0407m_e$, and $\gamma=2.46$ meV. In Fig. \ref{fig:ftir}(a), there is good agreement between theory and experiments. As expected, $R_{s}$ is reciprocal (dashed lines) and $R_p$ is nonreciprocal (solid lines). In Fig. \ref{fig:ftir}(a), the shaded areas represent the standard deviation of $N=16$ scans. In the case of $R_s$, there does not appear to be any statistically significant differences between the reflectance spectra as a function of $B$, whereas in the case of $R_p$, the reflectance spectra are well-separated with little to no overlap in the error bars.

To comprehensively understand the nonreciprocity of $R_p$, we define the nonreciprocal reflectance contrast:

\begin{equation} \label{eq:contrast}
\Delta R_{\ell} \equiv R_{\ell}(\omega,\theta,+B) - R_{\ell}(\omega,\theta,-B).
\end{equation}

\noindent Using Eq. (\ref{eq:rs}), it can be shown that $\Delta R_{s}=0$, regardless of the values of $\omega$, $\theta$, and $B$. We experimentally confirmed this and plot the experimental values of $\Delta R_s$ with error bars in Fig. \ref{fig:spolnonrecip} in the Supplementary Material. Figure \ref{fig:ftir}(b) shows (i) experimental and (ii) theoretical values of $\Delta R_p$. Again, there is good agreement between theory and experiments; even small features such as the dip below zero after the peak in $\Delta R_p$ seem to appear in the experimental values (just below 80 meV). There are two notable trends in the data: (1) $\Delta R_p$ increases with increasing $B$, as expected, and (2) $\Delta R_p$ decreases with increasing $\theta$. Prior work has suggested that nonreciprocity should be stronger at grazing angles of incidence, e.g., due to coupling to nonreciprocal surface plasmon polaritons \cite{zhao2020,tsurimaki2020}. However, our results suggest that, at least in the absence of couplers such as gratings and prisms, nonreciprocity is stronger at lower angles of incidence. This is corroborated by the results shown in Fig. \ref{fig:ellips}: compared to $\theta=60\degree$, $\Psi$ and $\Delta$ change much more drastically as a function of $B$ at $\theta=45\degree$. However, it should be pointed out that at normal incidence ($\theta=0\degree$), $\Delta R_p=0$ because of symmetry, meaning that there exists a $\theta$ value that maximizes $\Delta R_p$. For the first and second samples, we theoretically predict that these $\theta$ values are 20.1\degree and 28.7\degree, respectively.

Our results show that statistically significant nonreciprocity can be experimentally observed even at low magnetic fields. In this case, we measured a maximum nonreciprocal reflectance contrast as low as $\sim2\%$ at $0.07$ T and as high as $\sim10\%$ at $0.16$ T. As discussed in Sec. \ref{sec:ellips}, the ability to measure changes in $R_p$ in response to small changes in $B$ points toward applications in remote magnetic field sensing. In fact, measuring both $\Delta R_p$ and $\Delta R_s$ can have advantages over measuring $\Psi$ and $\Delta$. For example, the maximum nonreciprocal reflectance contrast does not seem to red- or blueshift as a function of $B$, which means they can be measured at a single appropriately chosen wavelength to sense $B$. (This may not be true at high magnetic fields since $m^*$ can be dependent on $B$, which would limit the range of $B$ values that can be sensed.) In addition, since $\Delta R_s$ is always zero, it can be used for calibration or as a litmus test of the fidelity of the measurement.

\section{\label{sec:conc}Conclusion}

\begin{figure}
\centering\includegraphics[scale=0.75]{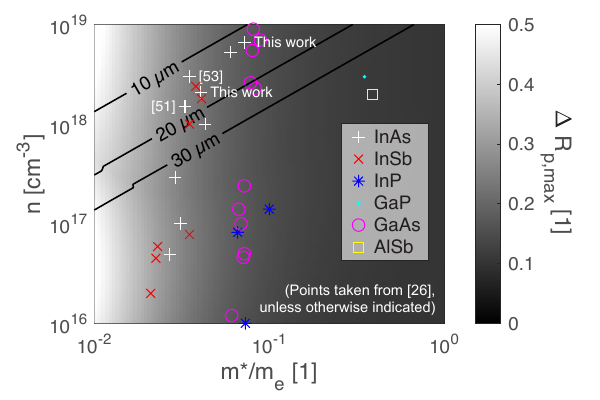}
\caption{Maximum nonreciprocal reflectance contrast $\Delta R_{p,\textrm{max}}\equiv\max_{\omega,\theta}{\Delta R_p}$ of a magneto-optic material modeled by the gyrotropic Drude-Lorentz model (Eqs. (\ref{eq:epsilon})--(\ref{eq:gyration})), plotted as a function of doping level $n$ and electronic effective mass $m^*$. It is assumed that $\epsilon_{\infty}=12$, $\gamma=0.1$ meV, and $B=0.1$ T, which are typical values. The different colored points are taken from \cite{palik1967}, unless otherwise indicated. White crosses are InAs; red X's InSb; blue asterisks InP; cyan points GaP; magnenta circles GaAs, and yellow squares AlSb. The contour lines are are the wavelengths at which $\Delta R_{p,\textrm{max}}$ is found for given values of $m^*$ and $n$.}
\label{fig:semicond}
\end{figure}

In this work, we experimentally observed nonreciprocal reflection of mid-IR light by highly doped InAs at low magnetic fields using two experimental techniques: ellipsometry and FTIR. We showed that $\Psi$ and $\Delta$ as well as $R_p$ and $R_s$ are nonreciprocal, i.e., they are directionally asymmetric with respect to $\theta$ or, equivalently, $B$. Our experiments showed good agreement with theory, experimentally validating the gyrotropic Drude-Lorentz model and showing that it is possible to fit the dielectric tensor using $\Psi$ and $\Delta$ (or even $R_p$ and $R_s$), without the need for Mueller matrix measurements. Our experiments demonstrated that $\Delta R_p$ can be large enough to be measured at magnetic fields as low as 0.07 T without the need for couplers such as gratings and prisms, suggesting that nonreciprocal thermal emitters and absorbers can be simpler in design. Our results suggest that spectra measured using both ellipsometry and FTIR can be sensitive to the magnitude and direction of the external magnetic field, pointing toward the possibility of remote magnetic field sensing using mid-IR light.

Although InAs has been the most popular platform for mid-IR nonreciprocity, there are other highly doped semiconductors that are known to host magnetoplasmons and are underexplored \cite{palik1967,zvezdinkotov}. To illustrate this parameter space, we plot $\Delta R_{p,\textrm{max}}\equiv \max_{\omega,\theta}\Delta R_p$ as a function of $n$ and $m^*$, assuming $\epsilon_{b}=12$, $\gamma=1$ meV, and $B=0.1$ T. This is shown in Fig. \ref{fig:semicond}. The different colored points in Fig. \ref{fig:semicond} are taken from \cite{palik1967} (unless otherwise indicated) and represent different III-V semiconductors in the same family as InAs. As can be seen, $n$ and $m^*$ span a wide range of values, but InAs, InSb, and GaAs seem the most promising since their maximum values of $\Delta R_p$ are relatively high and lie in the mid-IR, indicated by the contour lines in black. It also appears that $m^*$ is the more decisive factor that determines the strength of nonreciprocity, as $\Delta R_{p,\textrm{max}}$ does not seem to change much as a function of $n$. This makes sense, however, since the magnitude of the off-diagonal component of the dielectric tensor $g$, which is causes nonreciprocity, depends on ${m^{*}}^{-2}$ and $n^1$ (see Eq. (\ref{eq:gyration})). In spite of this, the importance of $n$ cannot be understated because it sets the ENZ frequencies. In Fig. \ref{fig:semicond}, $n$ can take a much wider range of values than $m^{*}$, but only some $n$ values are predicted to support mid-IR nonreciprocity ($5\ \mu\textrm{m} \lesssim \lambda \lesssim 30\ \mu\textrm{m}$, where $\lambda$ is the wavelength). This is highlighted by the contour lines. For example, measurements of $\Psi$, $\Delta$, $R_p$, and $R_s$ for a sample of InSb (MSE Supplies LLC) that does not have a high enough $n$ can be found in Figs. \ref{fig:insbellips} and \ref{fig:insbrefl} in the Supplementary Material. Despite being magnetized by a 0.12 T magnetic field, the optical response in the mid-IR is effectively reciprocal because ENZ frequencies are far away. With this in mind, the main conclusions of Fig. \ref{fig:semicond} are that $m^{*}$ should be as low as possible (to maximize nonreciprocity) and $n$ should be as high as possible (to set the ENZ frequencies in the mid-IR).

Further analysis in which $n$ is fixed and $\gamma$ is variable is shown in Fig. \ref{fig:drudelorentz} in the Supplementary Material and generally indicates that some loss can increase $\Delta R_p$ up to a point and that too little loss can decrease $\Delta R_p$. In fact, it can be proven that in the limit as $\gamma \rightarrow 0$, $\Delta R_{p} \rightarrow 0$, which implies that nonreciprocal reflection is consequence of nonreciprocal absorption, which is the fundamental physical process at work. 

It is worth noting that in Fig. \ref{fig:semicond} and throughout this work, we assumed that $m^{*}$ is independent of $\omega$, $n$, and $B$ when in reality, this may not be the case \cite{palik1967,wei2017}. The points in Fig. \ref{fig:semicond} were all measured at different values of $\omega$ and $B$ (although they were measured at room temperature). Assuming $m^*$ is constant seems to be a reasonable assumption given the good agreement between theory and experiments in this work and prior work. In the future, a better understanding of the relationships between the gyrotropic Drude-Lorentz model parameters could be extremely valuable, particularly in the context of nonreciprocity.

Going forward, although magnetic Weyl semimetals seem like the ideal solution for mid-IR nonreciprocity, they can be difficult to source. Highly doped semiconductors and nanofabrication with them are mature technologies in comparison, so future work in the field of nonreciprocal thermal radiation could benefit from exploring their potential. This includes measurements of their dielectric tensors and nonreciprocal reflectance contrast as a function of magnetic field. Experimental demonstrations of nanophotonically-enhanced nonreciprocal thermal radiation (beyond what has already been achieved) would be valuable since in principle, the spectral, directional, and polarization characteristics of thermal radiation can be fully controlled using nanophotonics. For example, since \textit{s}-polarized light is reciprocal, it can have a parasitic effect on nonreciprocity since on average, real-world thermal emitters equally emit both polarizations of light. Therefore, even if $\Delta R_{p}=1$, if $R_{s}=1$, then the polarization averaged nonreciprocal reflection contrast is only 0.5. Using, e.g., ENZ modes to suppress \textit{s}-polarized emission and absorption would help alleviate this.

\section*{Funding}
S. P., Y. T., and S. V. B. acknowledge support from ARO MURI (Grant
No. W911NF-19-1-0279) via U. Michigan. S. P. gratefully acknowledges support from the NSF GRFP under Grant No. 2141064.

\section*{Acknowledgments}
The authors thank Arun Nagpal, Qiuyuan Wang, Taqiyyah Safi, Luqiao Liu, Thanh Nguyen, Manasi Mandal, Mingda Li, Steven Kooi, and Caroline Ross for helpful discussions about magnetic materials, magneto-optic effects, and nonreciprocity over the course of this project. The authors thank Tim McClure and Charlie Settens for training and help with Fourier-transform infrared spectroscopy; Kevin Grossklaus and John McElearney for training and help with ellipsometry and Hall effect measurements; David Bono and Brian Neltner for their advice on measuring magnetic fields; Donal Jamieson and Kurt Broderick for training at MIT.nano; Shaymus Hudson for training and help with polishing; Duo Xu for help with 3D printing and measuring magnetic field; and Remi Sandell for help with design and prototyping during the early stages of this project.
This work was carried out in part through the use of MIT.nano's facilities. Ellipsometry was performed at the Tufts Epitaxial Core Facility using equipment supported by the United States Office of Naval Research (No. ONR DURIP N00014-17-1-2591). This material is based upon work sponsored (in part) by the U.S. Army DEVCOM ARL Army Research Office through the MIT Institute for Soldier Nanotechnologies under Cooperative Agreement number W911NF-23-2-0121.

\section*{Disclosures}
The authors declare no conflicts of interest.

\section*{Data Availability Statement}
Data underlying the results presented in this paper are not publicly available at this time but may be obtained from the authors upon reasonable request. CAD models of the magnetization schemes shown in Fig. \ref{fig:setup} may be obtained from the authors upon request.

\nocite{*}

\bibliographystyle{ieeetr}
\bibliography{references}

\widetext
\begin{center}
\textbf{\large Supplementary Material}
\end{center}
\setcounter{equation}{0}
\setcounter{figure}{0}
\setcounter{table}{0}
\setcounter{section}{0}
\makeatletter
\renewcommand{\theequation}{S\arabic{equation}}
\renewcommand{\thefigure}{S\arabic{figure}}

\section{Magnetization schemes}

\begin{figure}[h!]
    \centering
    \includegraphics[scale=0.65]{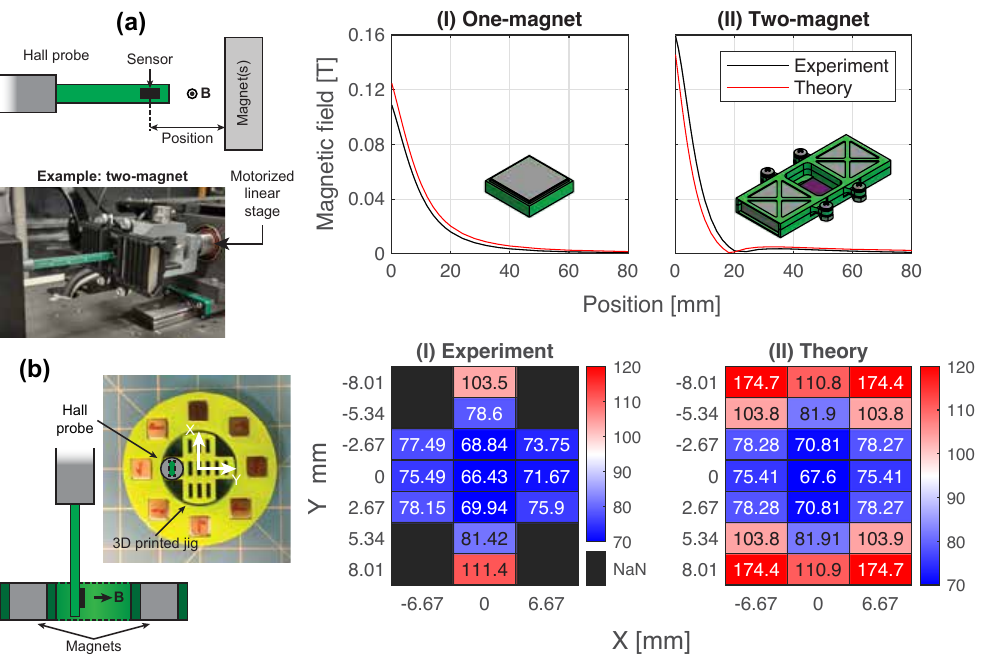}
    \caption{Measurement of the magnetic fields generated by the magnetization schemes. (a) One-magnet (I) and two-magnet (II) measured using a motorized linear stage, as shown in the schematic and photograph. Plots of the experimental (black) and theoretical values (red) are shown. The error bars associated with the experimental values are too small to be easily visible. (b) Halbach array measured using a 3D printed jig, as shown in the schematic and photograph. Heatmaps of experimental (I) and theoretical (II) values are shown. COMSOL Multiphysics\textregistered\ was used to numerically simulate the experimental setups (see Fig. \ref{fig:setup} in the main text for the magnetic field distributions generated by each one).}
    \label{fig:magnetization}
\end{figure}

Figure \ref{fig:magnetization} shows how we measured the magnetic fields generated by the magnetization schemes shown in Fig. \ref{fig:setup}. ``One-magnet'' and ``two-magnet'' were measured using a linear motorized stage. Each magnetization scheme was held in place using a clamp while a Hall probe was fixed at the other end of the machine using a 3D printed holder, which was designed to be aligned with the center of where the InAs wafer would be located during ellipsometry and FTIR. This is shown in Fig. \ref{fig:magnetization}(a). The magnetic field was measured as a function of the transverse position of the Hall probe relative to the magnets, which was digitally controlled by the machine, and compared to calculations done using COMSOL Multiphysics\textregistered, as shown in Fig. \ref{fig:magnetization}(a)(I)--(II). ``Halbach array'' was measured using a 3D printed jig with slots to insert the Hall probe and measure the magnetic field as a function of the lateral positions X and Y, as shown in Fig. \ref{fig:magnetization}(b). Again, this was compared to COMSOL Multiphysics\textregistered\ calculations, Fig. \ref{fig:magnetization}(b)(I)--(II).

\newpage
\section{Additional ellipsometry data for InAs}

\begin{figure}[h!]
    \centering
    \includegraphics[scale=0.6]{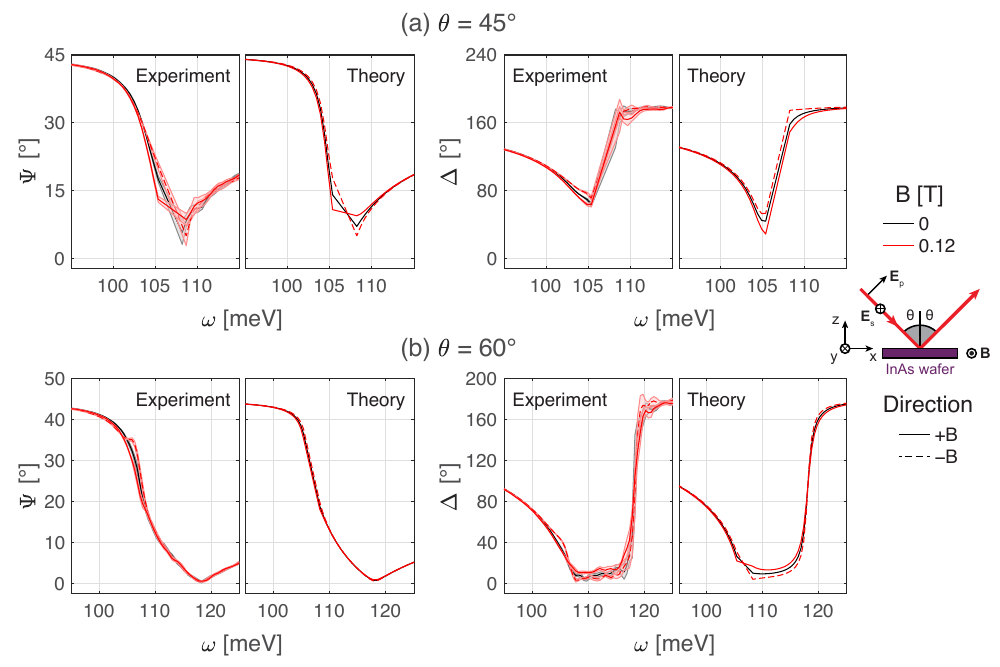}
    \caption{$\Psi$ (left two columns) and $\Delta$ (right two columns) measured using ellipsometry at $B$ values of 0 T (black, for reference) and 0.12 T (red). (a) $\theta=45\degree$. Experimental and theoretical values are indicated. Solid and dashed lines correspond to $B$ pointing in the $+y$- or $-y$-direction, respectively. The shaded areas in the experimental values are the standard deviation ($N=50$). (b) Same as (a), but $\theta=60\degree$. The inset schematic shows the coordinate system, the $B$ vector, $\theta$, and \textit{s}- and \textit{p}-polarized electric field vectors.}
    \label{fig:addellips}
\end{figure}

\begin{figure}[h!]
    \centering
    \includegraphics[scale=0.6]{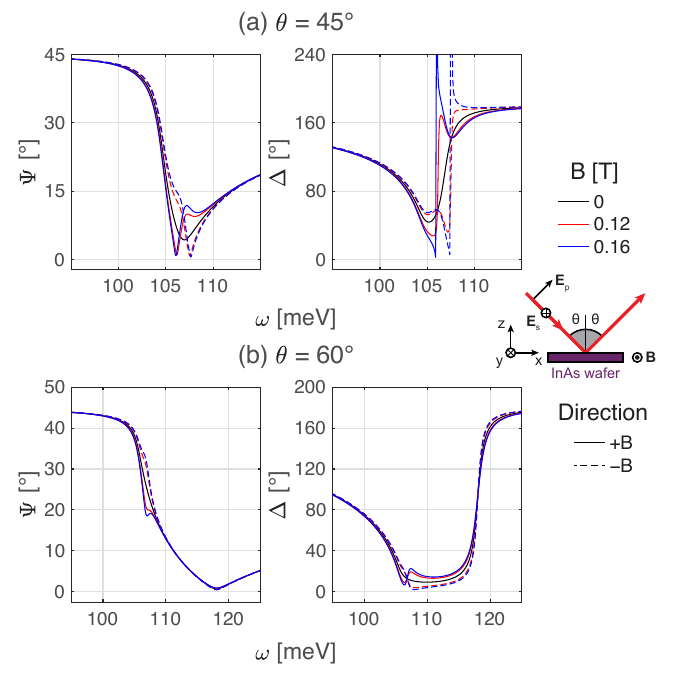}
    \caption{Theoretical $\Psi$ and $\Delta$ at $B$ values of 0 T (black), 0.12 T (red), and 0.16 T (blue), calculated using a finer spectral resolution than in the main text. Sharp features are present that are not apparent in the experimental values. (a) $\theta=45\degree$. Solid and dashed lines correspond to $B$ pointing in the $+y$- or $-y$-direction, respectively. (b) Same as (a), but $\theta=60\degree$. The inset schematic shows the coordinate system, the $B$ vector, $\theta$, and \textit{s}- and \textit{p}-polarized electric field vectors.}
    \label{fig:theoellips}
\end{figure}

This section contains additional ellipsometry data for InAs. Figure \ref{fig:addellips} shows experimental $\Psi$ and $\Delta$ for the intermediate $B$ value of 0.12 T, not shown in the main text to prevent clutter. Experiments show relatively good agreement with theory and follow the predicted trend of increasing strength of nonreciprocity with increasing $B$. Figure \ref{fig:theoellips} shows the same theoretical values of $\Psi$ and $\Delta$ as in Fig. \ref{fig:ellips} but calculated using a finer spectral resolution. Sharp features whose frequency and bandwidth depend on the magnitude and direction of $\textbf{B}$ become apparent, which may be useful for remote magnetic field sensing.

\newpage
\section{Additional reflectance data for InAs}

\begin{figure}[h!]
    \centering
    \includegraphics[scale=0.8]{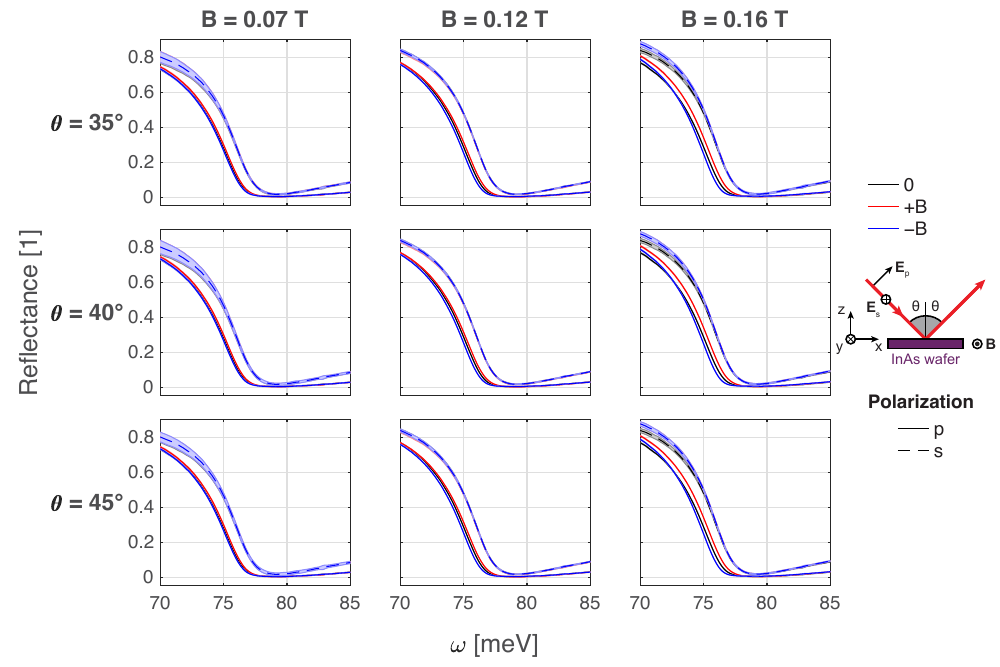}
    \caption{Experimental $R_{\ell}(\omega,\theta,B)$ ($\ell=s,p$) as a function of $\theta$, $B$, and $\ell$. The black lines correspond to $B=0$ T; red $B$ pointing in the $+y$-direction; and blue $B$ pointing in the $-y$-direction. Solid and dashed lines correspond to \textit{p}- and \textit{s}-polarization, respectively. The shaded areas in the experimental values are the standard deviation ($N=16$). The inset schematic shows the coordinate system, the $B$ vector, $\theta$, and \textit{s}- and \textit{p}-polarized electric field vectors.}
    \label{fig:addrefl}
\end{figure}

\begin{figure}[h!]
    \centering
    \includegraphics[scale=0.8]{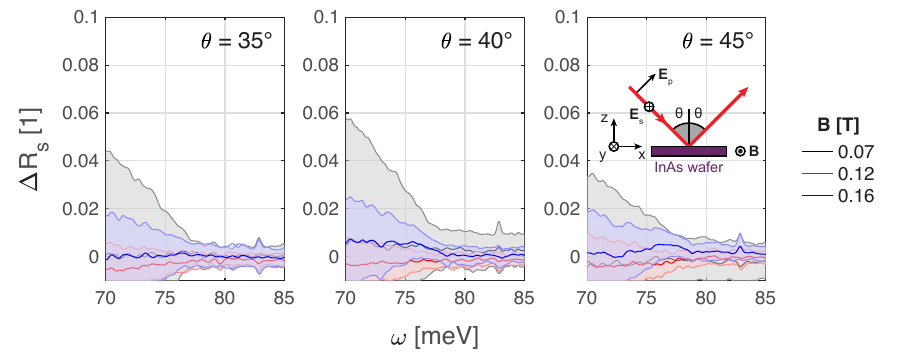}
    \caption{Experimental nonreciprocal reflectance contrast of \textit{s}-polarized light, $\Delta R_{s} \equiv R_{s}(\omega,\theta,+B)-R_{s}(\omega,\theta,-B)$. The black lines correspond to $B=0.07$ T; red 0.12 T; and blue 0.16 T. The shaded areas in the experimental values are the standard deviation ($N=16$). As expected, $\Delta R_{s}=0$. The vertical scale is identical to Fig. 4 in the main text to facilitate comparison. The inset schematic shows the coordinate system, the $B$ vector, $\theta$, and \textit{s}- and \textit{p}-polarized electric field vectors.}
    \label{fig:spolnonrecip}
\end{figure}

This section contains additional reflectance data for InAs not shown in the main text. Figure \ref{fig:addrefl} shows all the experimental $R_{\ell}(\omega,\theta,B)$ ($\ell=s,p$) we measured and used to calculate $\Delta R_{p}$ in Fig. \ref{fig:ftir}(b). Figure \ref{fig:spolnonrecip} shows experimental values of $\Delta R_{s}$, validating that it is zero as predicted by Eq. (\ref{eq:rs}).

\newpage
\section{Ellipsometry and reflectance data for InSb}

\begin{figure}[h!]
    \centering
    \includegraphics[scale=0.6]{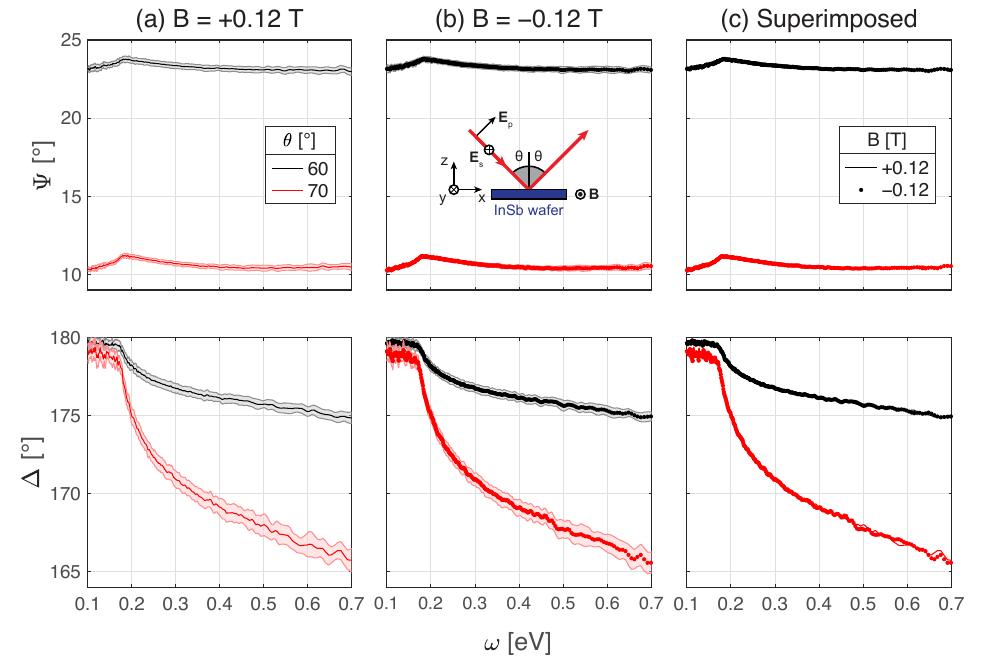}
    \caption{Experimental $\Psi$ and $\Delta$ for InSb at a B value of 0.12 T. The black and red lines correspond to $\theta=60\degree$ and $70\degree$, respectively. The shaded areas in the experimental values are the standard deviation ($N=40$). (a) Case of $B$ pointing in the $+y$-direction. (b) Case of $B$ pointing in the $-y$-direction. The inset schematic shows the coordinate system, the $B$ vector, $\theta$, and \textit{s}- and \textit{p}-polarized electric field vectors. (c) The plots in (a) and (b) superimposed on each other, without the standard deviation. The solid lines correspond to $B=+0.12$ T, and the dots correspond to $B=-0.12$ T. Although InSb can support free-carrier magneto-optic effects and nonreciprocal electromagnetic modes, the doping level is not high enough to do so in the mid-infrared. As a result, $\Psi$ and $\Delta$ are effectively reciprocal even though the sample was magnetized.}
    \label{fig:insbellips}
\end{figure}

\begin{figure}[h!]
    \centering
    \includegraphics[scale=0.6]{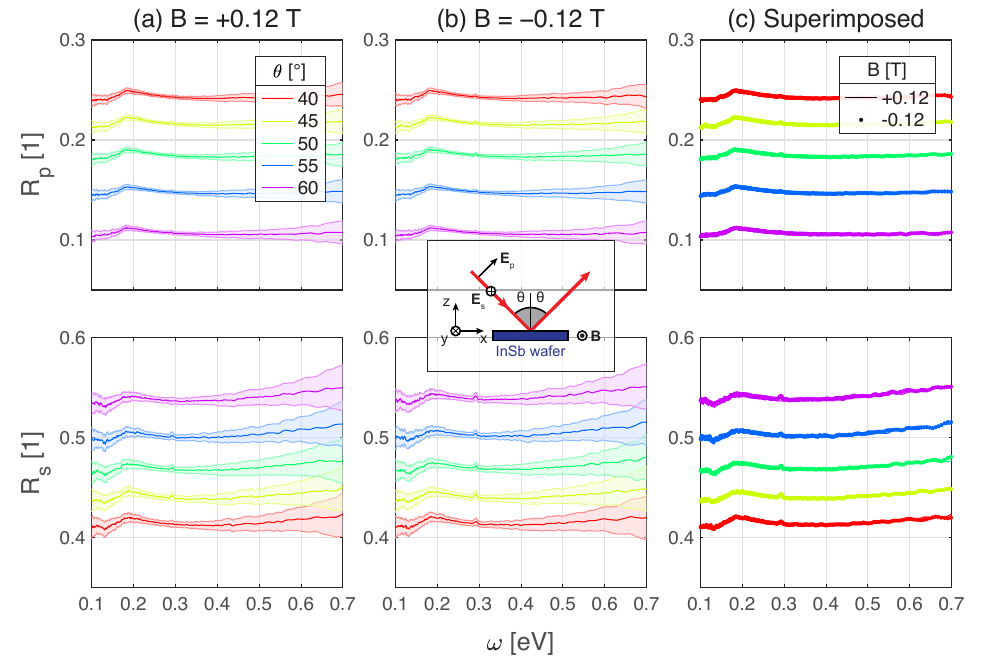}
    \caption{Experimental $R_p(\omega,\theta,B)$ and $R_s(\omega,\theta,B)$ for InSb at a B value of 0.12 T. The different colors correspond to different $\theta$ values (see the legend). The shaded areas in the experimental values are the standard deviation ($N=40$). (a) Case of $B$ pointing in the $+y$-direction. (b) Case of $B$ pointing in the $-y$-direction. The inset schematic shows the coordinate system, the $B$ vector, $\theta$, and \textit{s}- and \textit{p}-polarized electric field vectors. (c) The plots in (a) and (b) superimposed on each other, without the standard deviation. The solid lines correspond to $B=+0.12$ T, and the dots correspond to $B=-0.12$ T. Similar to Fig. \ref{fig:insbellips}, the optical response is effectively reciprocal even though the sample was magnetized.}
    \label{fig:insbrefl}
\end{figure}

This section contains ellipsometry and reflectance data for an InSb wafer that does not support nonreciprocity due to low doping. Figure \ref{fig:insbellips} shows experimental $\Psi$ and $\Delta$ for $B=\pm 0.12$ T, showing no statistically significant difference between them (meaning they are reciprocal). Similarly, Figure \ref{fig:insbrefl} shows $R_{p}(\omega,\theta,B)$ and $R_{s}(\omega,\theta,B)$, both of which are reciprocal. This illustrates that not just any highly doped semiconductor can support mid-IR nonreciprocity; appropriate choices of $n$ and $m^*$ are required.

\newpage
\section{Additional analysis of the gyrotropic Drude-Lorentz model}

\begin{figure}[h!]
    \centering
    \includegraphics[scale=0.75]{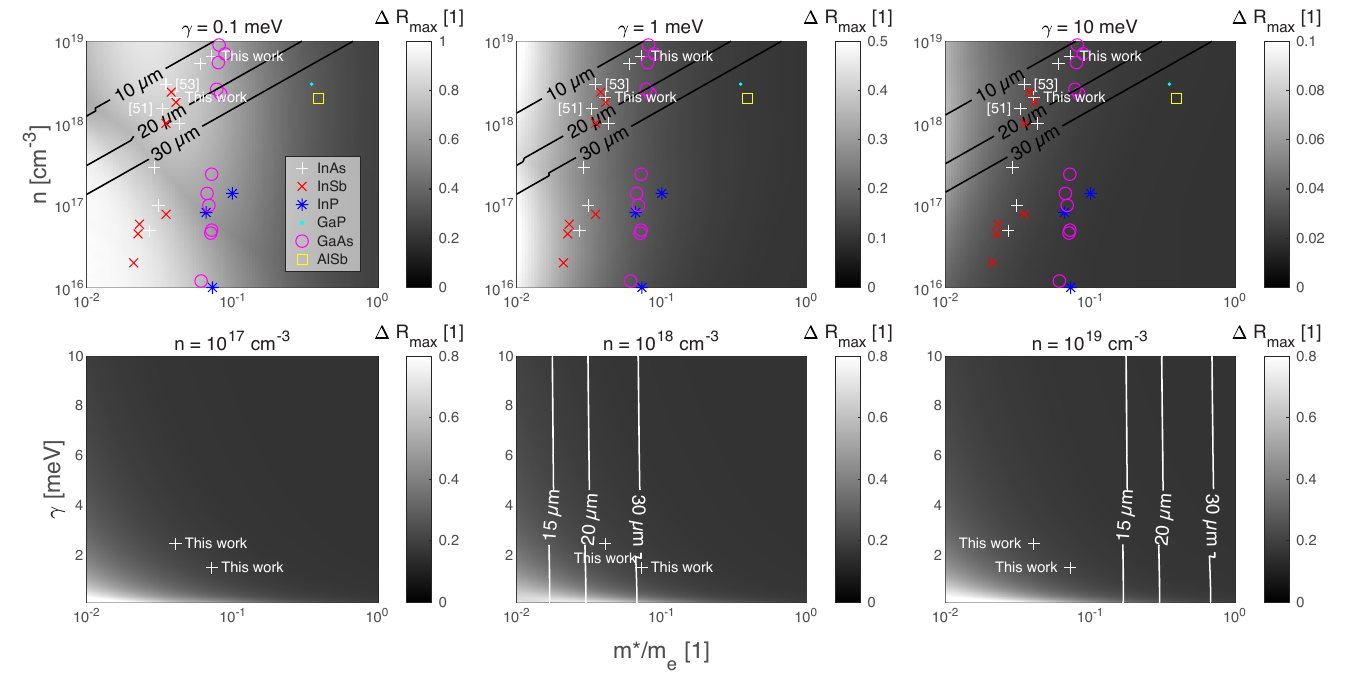}
    \caption{Maximum nonreciprocal reflectance contrast $\Delta R_{p,\textrm{max}}\equiv\max_{\omega,\theta}{\Delta R_p}$ of a magneto-optic material modeled by the gyrotropic Drude-Lorentz model (Eqs. (2)--(5) in the main text), plotted as a function of doping level $n$, electronic effective mass $m^*$, and damping constant $\gamma$. The horizontal axis is always $m^{*}$, while the vertical axis is either $n$ (while $\gamma$ is fixed, top row) or $\gamma$ (while $n$ is fixed, bottom row). It is assumed that $\epsilon_{\infty}=12$ and $B=0.1$ T, which are typical values, and the fixed value of $\gamma$ or $n$ is stated above each plot in the top or bottom row, respectively. The different colored points are taken from \cite{palik1967}, unless otherwise indicated. White crosses are InAs; red X's InSb; blue asterisks InP; cyan points GaP; magnenta circles GaAs, and yellow squares AlSb. In the bottom row, since damping constants are not provided in \cite{palik1967}, only the values from this work are shown (see the main text). The contour lines (black or white) are are the wavelengths at which $\Delta R_{p,\textrm{max}}$ is found for given values of $m^*$ and $n$ or $\gamma$.  When $n=10^{17}\ \textrm{cm}^{-3}$, the maximum value of $\Delta R_{p,\textrm{max}}$ is never in the mid-infrared, so there are no contour lines (i.e., this value of $n$ is unlikely to be useful for mid-infrared nonreciprocity unless $m^{*}$ is extremely low ($\ll 10^{-2}m_{e}$)).}
    \label{fig:drudelorentz}
\end{figure}

Additional analysis of the gyrotropic Drude-Lorentz model is shown in Fig. \ref{fig:drudelorentz}, allowing for arbitrary $n$, $m^*$, and $\gamma$, reveals a number of interesting trends. Firstly, it appears that $n$ does not strongly influence $\Delta R_{p,\textrm{max}}$ and instead sets the peak wavelength of nonreciprocity (black and white contour lines in the top and bottom rows of Fig. \ref{fig:drudelorentz}, respectively). Therefore, for a semiconductor to support mid-IR nonreciprocity (via free-carrier magneto-optic effects), $n$ needs to be sufficiently high. On the other hand, $m^*$ plays an important role in setting the magnitude of $\Delta R_{p,\textrm{max}}$, and in fact, it appears that the lower the value of $m^*$, the better, although as can be seen in the bottom row of $\ref{fig:drudelorentz}$, it also depends on $\gamma$. Generally, it appears that lower values of $\gamma$ lead to stronger nonreciprocity, but in the limit as $\gamma\rightarrow 0$, it can be proven using Eqs. (\ref{eq:transverse})--(\ref{eq:gyration}) and (\ref{eq:rp}) that $\Delta R_{p}=0$ because there is no absorption at all, and absorption is the fundamentally nonreciprocal process at play.

\end{document}